\documentclass[aps,twocolumn]{revtex4}
\usepackage{epsfig}
\begin{document}
\def\I.#1{\it #1}
\def\B.#1{{\bf #1}}
\def\C.#1{{\cal  #1}}

\title{Quasi-Static Brittle Fracture in Inhomogeneous Media and Iterated Conformal
Maps: Modes I, II and III}
\author{Felipe Barra$^*$, Anders Levermann and Itamar Procaccia}
\affiliation{Dept. of Chemical Physics, 
The Weizmann Institute of Science, Rehovot, 76100, Israel\\
$^*$ Present address:
Dept. Fisica, Facultad de Ciencias Fisicas y Matematicas, Universidad de
chile,
Casilla 487-3, Santiago, Chile.}
\widetext
\begin{abstract}
The method of iterated conformal maps is developed for quasi-static
fracture of brittle materials, for all modes of fracture. Previous
theory, that was relevant for mode III only, is extended here to
mode I and II. The latter require solution of the bi-Laplace rather
than the Laplace equation. For all cases we can consider quenched
randomness in the brittle material itself, as well as randomness in
the succession of fracture events. While mode III calls for the advance
(in time) of one analytic function, mode I and II call for the
advance of two analytic functions. This fundamental difference
creates different stress distribution around the cracks. As a result
the geometric characteristics of the cracks differ, putting mode III
in a different class compared to modes I and II.
\end{abstract}
\maketitle
\section{Introduction}

The theory of quasi-static fractures in brittle media \cite{86LL,52Mus,90HR,99FM,98Fre} 
calls for solving
different equations depending on the mode of fracture. In this paper we 
present an approach based on iterated conformal maps which can be
adapted to solve all three modes of fracture (known as mode I, II and III), 
including the effects
of inhomogeneities and randomness of the brittle material itself. 

Basically, the theory of fracture in brittle continuous media is based
on the equation of motion for an isotropic elastic body in the
continuum limit \cite{86LL}
\begin{equation} 
\rho \frac{\partial^2\B.u}{dt^2}=(\lambda+\mu)
\B.\nabla(\B.\nabla\cdot \B.u)+\mu \nabla^2\B.u \ . \label{eqmot}
\end{equation}
Here $\B.u$ is the field describing the displacement of each mass point from its location
in an unstrained body and $\rho$ is the density. 
The constants $\mu$ and $\lambda$ are the Lam\'e constants.
In terms of the displacement field the elastic strain tensor is defined as
\begin{equation}
\epsilon_{ij}\equiv \frac{1}{2}\left(\frac{\partial u_i}{\partial x_j}
+\frac{\partial u_j}{\partial x_i}\right) 
\ . \label{strain}
\end{equation}
For the development of a crack the important object is the stress tensor, 
which in linear elasticity is written as
\begin{equation}
\sigma_{ij}\equiv \lambda \delta_{ij}\sum_k\epsilon_{kk} +2 \mu \epsilon_{ij}
\ . \label{stress}
\end{equation}
When the stress component which is transverse to the interface of a crack exceeds a threshold value
$\sigma_c$, the crack can develop. When the external load is such that the transverse stress exceeds
only slightly the threshold value, the crack develops slowly, and one can neglect the second 
time-derivative in Eq. (\ref{eqmot}). This is the quasi-static limit, in which after each growth event
one needs to recalculate the strain field by solving the Lam\'e equation
\begin{equation}
(\lambda+\mu)\B.\nabla(\B.\nabla\cdot \B.u)+\mu \nabla^2\B.u =0 \ . \label{lame}
\end{equation}
The three ``pure" modes of fracture that can be considered are determined 
by the boundary conditions, or load, at infinity. These are
\begin{eqnarray}
&&\!\!\!\!\!\!\!\!\sigma_{xx}(\infty)=0\ ; \sigma_{yy}(\infty)=\sigma_\infty\ ; 
 \sigma_{xy}(\infty)=0\quad \text{Mode I}\label{mode1}\\
&&\!\!\!\!\!\!\!\!\sigma_{xx}(\infty)=0\ ; \sigma_{yy}(\infty)=0\ ; 
 \sigma_{xy}(\infty)=\sigma_\infty\quad \text{Mode II} \ . \label{mode2}
\end{eqnarray}
We will study the fracture patterns of these two modes in 2-dimensional
materials.
Mode III calls for a third dimension $z$, since 
\begin{equation}
\sigma_{zy}(y\to \pm \infty)= \sigma_\infty \quad \text{Mode III} . \label{mode3}
\end{equation}
Such an applied stress creates a displacement field $u_z(x,y)$, $u_x=0$,
$u_y=0$ in the medium. Thus, in spite of the third dimension, 
the calculation of the strain
and stress tensors remain two dimensional. Nevertheless, the equations
to be solved in mode III and modes I and II are different.
In mode III fracture $\B.\nabla \cdot \B.u=0$, and the Lam\'e equation
reduces to Laplace's equation
\begin{equation}
\label{laplace}
\Delta u_z\equiv \partial^2 u_z/\partial x^2 + \partial^2 u_z/\partial y^2 = 0,
\end{equation}
and therefore $u_z$ is
the real part, Re $\chi(z)$, of an analytic function $\chi(z)$,
\begin{equation}
\label{analytic}
\chi (z) = u_z(x,y) + i \xi_z(x,y) \ ,
\end{equation}
where $z = x+iy$. The boundary conditions far from the crack and on the crack interface can be used
to find this analytic function. On the other hand, for mode I and mode II fractures in
plane elasticity one introduces \cite{86LL} the Airy potential $U(x,y)$ such that
\begin{equation}
\sigma_{xx}=\frac{\partial^2 U}{\partial y^2} \ ; \sigma_{xy}=- 
\frac{\partial^2 U}{\partial x\partial y}
\ ; \sigma_{yy}=\frac{\partial^2 U}{\partial x^2} \ . \label{sigU}
\end{equation}
The Airy potential $U$ solves the bi-Laplacian equation \cite{52Mus}
\begin{equation}
\Delta \Delta U(x,y) =0 \ . \label{bilaplace}
\end{equation}
The solution of the bi-Laplacian equation can be written  in terms
of {\em two} analytic functions $\phi(z)$ and $\eta(z)$ as
\begin{equation}
U(x,y)= \text{Re} [\bar z\varphi(z)+\eta(z)] \ . \label{Uphichi}
\end{equation}
This difference requires therefore a separate discussion of mode
III and modes I and II. 

The problem of quasi-static crack propagation is difficult not only because it is hard
to solve Eq. (\ref{lame}) for an arbitrarily shaped crack. Another source
of difficulty is that the equation does not dictate how to propagate a 
crack when the stress tensor exceeds the threshold value $\sigma_c$.
In this paper we consider only 2-dimensional, or effectively
2-dimensional (i.e thin slabs) brittle materials in $x,y$.  
We can then describe a crack of arbitrary shape by its interface $\vec{x}(s)$,
where $s$ is the arc length which is used to parameterize the contour. 
We will use the notation $(t,n)$ to describe
respectively the transverse and normal directions at any point on the two-dimensional crack interface.
The literature is quite in agreement that the velocity of propagation
of the crack has a normal component which is some function of $\sigma_{tt}(s) - \sigma_c$
for mode I and II, and of $|\sigma_{zt}(s)| - \sigma_c$ for mode III. In both cases
$\sigma_c$ is a measure of the strength of the material, and fracture occurs only
if the local stress tensor at the boundary of the crack exceeds this
quantity (which can also be a random function of position).
There is hardly a consensus however on what that function is.
The simplest choice \cite{89BDL,90Ker} is a linear function, 
\begin{eqnarray}
v_n(s) &=&\alpha\Delta \sigma\equiv \alpha (\sigma_{tt}(s) - \sigma_c(s)) \ , \quad \text{Mode I,II} \ ,
\label{vel}
\\ v_n(s) &=&\alpha\Delta \sigma\equiv \alpha (|\sigma_{zt}(s)| - \sigma_c(s)) \ , \quad \text{mode III}
\ .
\label{velocity}
\end{eqnarray}
when $\Delta \sigma\ge 0$, and $v_n(s)=0$ otherwise. 
Other velocity laws are possible \cite{2.3}. In our study of mode III fracture we will examine also a
quadratic and an exponential velocity law:
\begin{eqnarray}
v_n(s) &=& \alpha (|\sigma_{zt}(s)| - \sigma_c(s))^2 \ , \quad \text{mode III} \ ,\label{quad}\\
v_n(s) &=&e^{ \alpha (|\sigma_{zt}(s)| - \sigma_c(s))} \ , \quad \text{mode III} \ .  \label{exp}
\end{eqnarray}
It is important to study these variants of the velocity law to ascertain the degree
of universality of the geometric characteristics of the resulting cracks. 
One of our results is that these characteristics {\em may depend} on the
velocity law. While this may be a disappointment from the point of view of fundamental
physics, it may help to identify the correct physical mechanisms of fractures in 
different media. The lack of universality is even more obvious when we add quenched noise,
or random values of $\sigma_c(s)$. The geometric characteristics of the cracks may depend on the
probability distribution of random values of $\sigma_c(s)$. Again this may give a handle on the
characterization of inhomogeneous brittle materials.

At any point
in time there can be more than one position $s$ on the interface for which $v_n(s)$ does not vanish. We 
choose the next growth position randomly with a probability proportional to $v_n(s)$ \cite{90Ker,87LG}.
There we extend the crack by a fixed area of the size of the ``process zone" (and see below for details).
This is similar to Diffusion Limited Aggregation (DLA) in which a particle is grown with a probability proportional
to the gradient of the field. One should note that another model could be derived in which all eligible fracture
sites are grown simultaneously, growing a whole layer whose local width is $v_n(s)$. This would be more akin to
Laplacian growth algorithms, which in general give rise to clusters in a different universality class than
DLA \cite{01BDLP,01BDP}.

In Sect. \ref{confsec} we discuss the growth algorithm in terms of iterated conformal
maps. In Sect. \ref{modeIII} this method is applied to mode III quasi-static
fracture. A preliminary report of the method for this case was presented
in \cite{02BHLP}. In Sect. IV we present new results including the consequences of
the different velocity laws (\ref{quad}) and (\ref{exp}), and those
of quenched randomness. We discuss the geometric properties of the fracture
patterns, including issues of roughening and exponents. We point out that the
roughening exponents are not always well defined, since the fracture patterns
do not have stationary geometric characteristics. There is an increased tendency for ramification
as the fracture develops. This is reflected in an apparent increase in the
roughening exponents of the backbone of the pattern. In Sect. \ref{modeI} we
discuss the theory of mode I and II fracture. Sect.\ref{mode12results} presents
the results. We will see that the fracture patterns in mode I and II are much less
rough then in mode III (for the same velocity law), in agreement with the 
analysis of \cite{97REF}. We will conclude the paper in Sect. \ref{conclusions}.
The main conclusion is that mode III results in cracks whose geometric 
characteristics are in a different class than modes I and II. The
former creates cracks that exhibit a cross over in the roughening exponent
from about 0.5 to a higher scaling exponent on the larger scales.
In contrast, modes I and II create cracks that are not rough on the
large scales. Quenched randomness may affect the geometry of the cracks as
is exemplified and discussed in this paper.
\section{The method of iterated conformal maps for fracture}
\label{confsec}
The direct determination of the strain tensor for an arbitrary shaped (and evolving) crack is
difficult. We therefore proceed by turning to a mathematical complex plane $\omega$,
in which the crack is forever circular and of unit radius. 
Next invoke a conformal map $z = \Phi^{(n)}(\omega )$ that maps the exterior of 
the unit circle in the mathematical plane
$\omega$ to the exterior of the crack in the physical plane $z$, after $n$ growth steps. The
conformal map will be univalent by construction, and we can write its
Laurent expansion in the form
\begin{equation}
\Phi^{(n)}(\omega ) = F_1^{(n)}\omega + F_0^{(n)} 
+F_{-1}^{(n)}/\omega+F_{-2}^{(n)}/\omega^2+\cdots \ . \label{Laurent}
\end{equation}
For all modes of fracture we take $\Phi^{(0)}(\omega )=\omega$, and the iterative
dynamics calls for the calculation of the transverse component
of the stress tensor on the boundary of the crack. The arclength position
$s$ in the physical domain is mapped by the inverse of $\Phi^{(n)}$ onto
a position the unit circle $\omega=\exp(i\theta)$.
We will be able to compute the stress tensor on the boundary of 
the crack in the physical domain by 
performing the calculation on the unit circle. In other words we
will compute $\sigma_{tt}(\theta)$ or
$\sigma_{zt}(\theta)$  on the unit circle in the mathematical plane. The actual calculation of
this component of the stress tensor differs in modes I,II and mode III.
We perform the calculation iteratively, taking the stress as known for
the crack after $n-1$ fracture events.

In order
to implement the $n$th cracking event according to one of the required velocity
laws (\ref{vel})-(\ref{exp}), 
we should choose potential positions on the interface
more often when $\Delta\sigma(\theta)$ is larger. Consider for
example the linear velocity law (\ref{vel}). We construct
a probability density $P(\theta)$ on the unit circle $e^{i\theta}$
which satisfies
\begin{equation}
P(\theta) = \frac{|\Phi^{'(n-1)}(e^{i \theta} )|\Delta\sigma(\theta)
\Theta(\Delta\sigma(\theta))}{\int_0^{2\pi}|\Phi^{'(n-1)}(e^{i \tilde\theta} )|\Delta\sigma(\tilde\theta)
\Theta(\Delta\sigma(\tilde\theta))d\tilde\theta}
\ , \label{weight}
\end{equation} 
where $\Theta(\Delta\sigma(\tilde\theta))$ is the Heaviside function, 
and $|\Phi^{'(n-1)}(e^{i \theta} )|$ is simply the Jacobian of the 
transformation from mathematical to physical plane. The next
growth position, $\theta_n$ in the mathematical plane, is chosen randomly with
respect to the probability $P(\theta) d\theta$. 
At the chosen position on the crack, i.e. $z= \Phi^{(n-1)}(e^{i\theta_n})$, we want to advance
the crack with a region whose area is the typical process zone for the material that we analyze.
According to \cite{90HR} the typical scale of the process zone is $K^2/\sigma^2_c$,
where $K$ is a characteristic fracture toughness parameter. Denoting the typical {\em area}
of the process zone by $\lambda_0$, we achieve growth with an auxiliary conformal map
$\phi_{\lambda_n,\theta_n}(\omega )$ that maps the unit circle to a unit circle with a bump
of area $\lambda_n$ centered at $e^{i\theta_n}$. 
An example of such a map is given by  \cite{98HL,99DHOPSS}:
\begin{eqnarray}
\label{phi}
   &&\phi_{\lambda,0}(w) = w \left\{ \frac{(1+
   \lambda)}{2w}(1+w)\right. \\
   &&\left.\times \left [ 1+w+w \left( 1+\frac{1}{w^2} -\frac{2}{w}
\frac{1-\lambda} {1+ \lambda} \right) ^{1/2} \right] -1 \right \} ^a \nonumber\\
   &&\phi_{\lambda,\theta} (w) = e^{i \theta} \phi_{\lambda,0}(e^{-i
   \theta}
   w) \,,
   \label{eq-f}
\end{eqnarray}
Here the bump has an aspect ratio $a$, $0\le a\le1$. 
In our work below we use $a=1/2$.
 To ensure a fixed size step in the physical domain we choose
\begin{equation}
   \lambda_{n} = \frac{\lambda_0}{|{\Phi^{(n-1)}}' (e^{i \theta_n})|^2} \ .
   \label{lambdan}
\end{equation}
Finally the updated conformal map $\Phi^{(n)}$ is obtained as
\begin{equation}
\label{conformal}
\Phi^{(n)}(\omega ) = \Phi^{(n-1)}(\phi_{\lambda_n,\theta_n}(\omega )) \ . \label{iter}
\end{equation} 

The recursive dynamics can be represented as iterations
of the map $\phi_{\lambda_{n},\theta_{n}}(w)$,
\begin{equation}
   \Phi^{(n)}(w) =
\phi_{\lambda_1,\theta_{1}}\circ\phi_{\lambda_2,\theta_{2}}\circ\dots\circ
\phi_{\lambda_n,\theta_{n}}(\omega)\ . \label{comp}
\end{equation} 
Every given fracture pattern is determined completely by the random itinerary
$\{\theta_i\}_{i=1}^n$. 
\begin{figure}
\centering
\includegraphics[width=.3\textwidth]{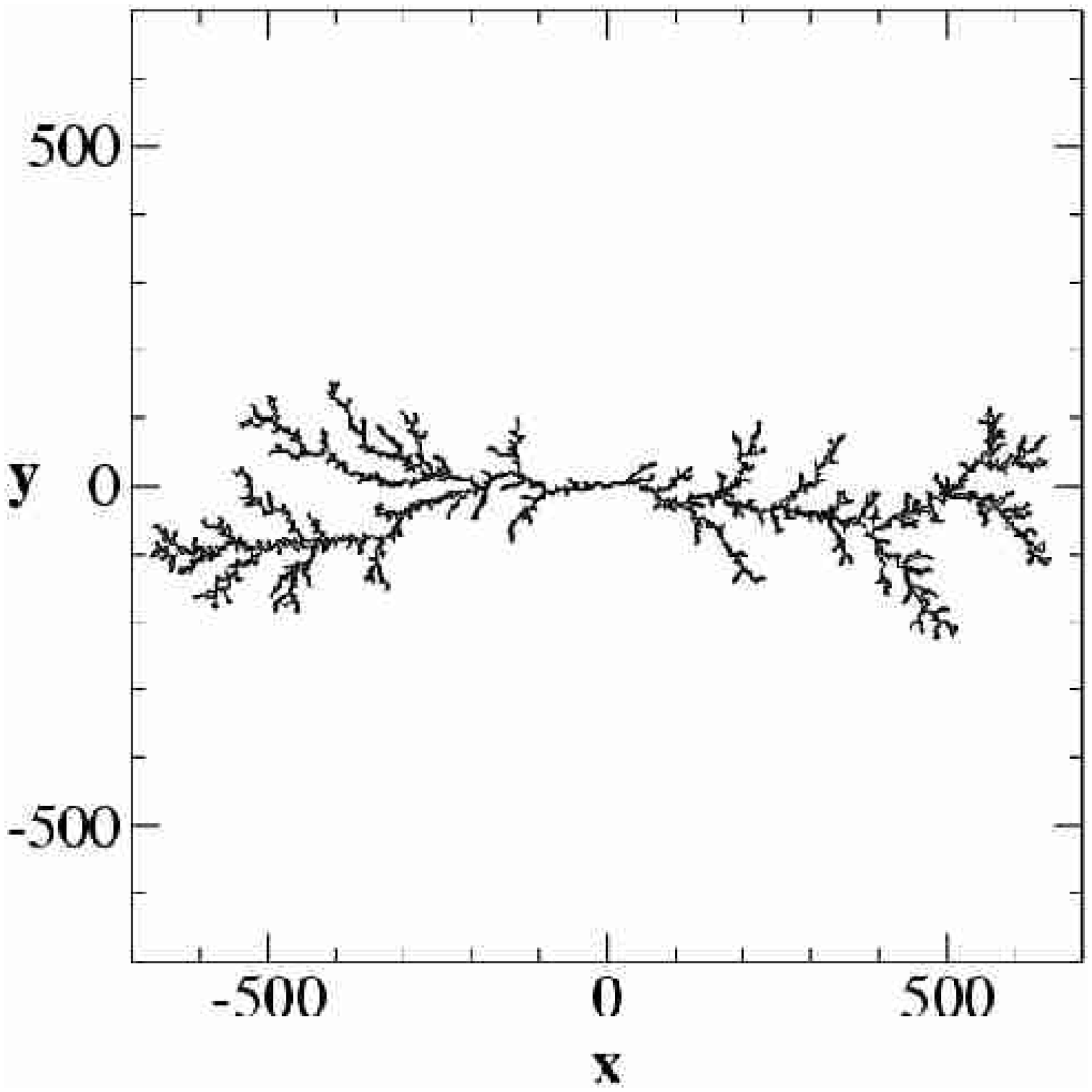}
\centering
\includegraphics[width=.3\textwidth]{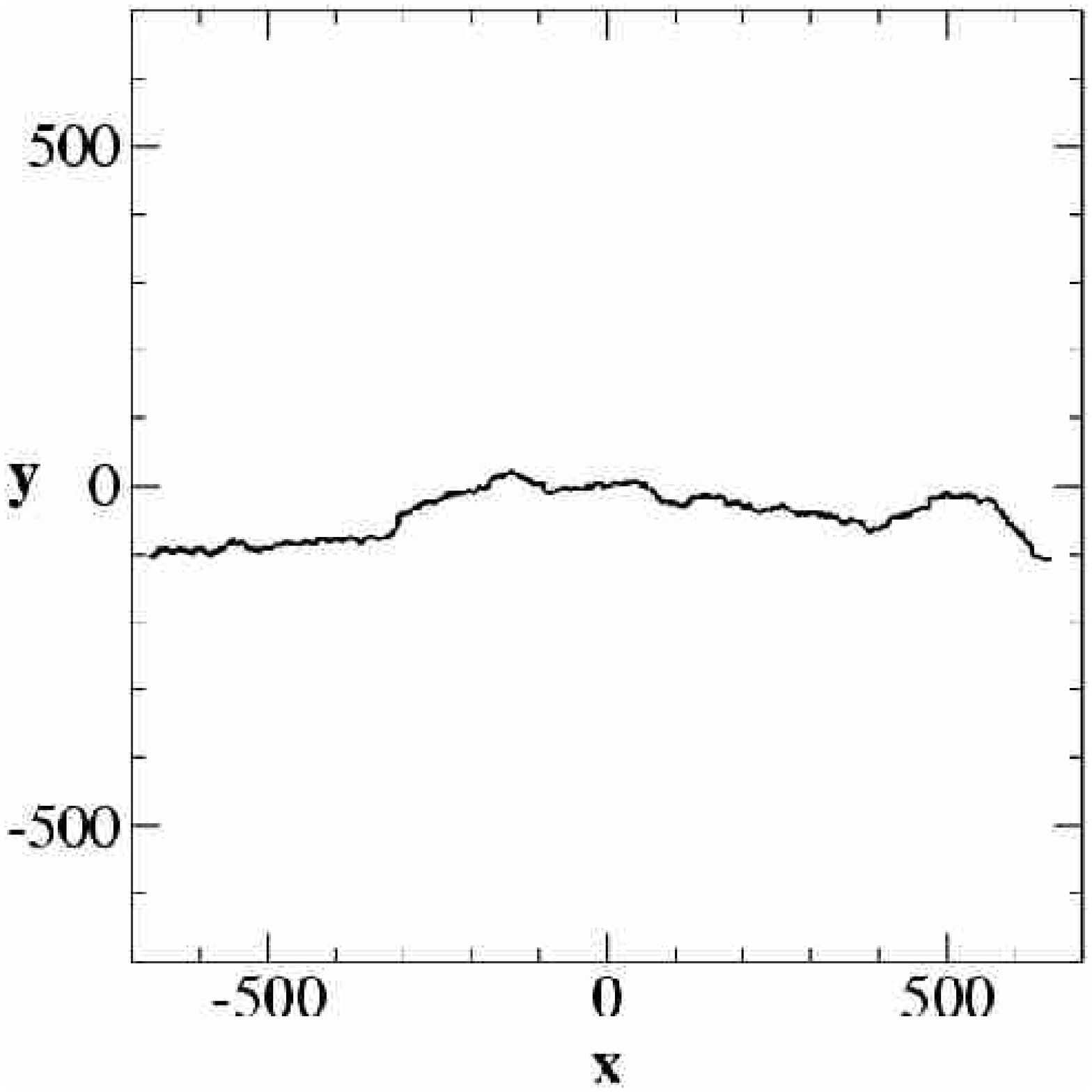}
\caption{Upper panel: a typical mode III fracture pattern that is obtained from
iterated conformal maps. What is seen is the boundary of the fractured zone, which is the
mapping of the unit circle in the mathematical domain onto the physical domain.
Notice that the pattern becomes more and more ramified as the the
fracture pattern develops. This is due to the enhancement of the stress field
at the tips of the growing pattern.
Lower panel: the backbone of the fracture pattern. This is the 
projection onto the x-y plane of the experimentally
observed boundary between the two parts of the material that separate
when the fracture pattern hits the lateral boundaries.}
\label{fracture}
\end{figure} 
\section{Mode III quasi-static fracture}
\label{modeIII}

In this section we discuss how to compute the stress tensor when the load
is mode III, using the method of iterated conformal maps. 
The first step is the determination of the boundary conditions that
the analytic function  (\ref{analytic}) needs to satisfy.
\subsection{Boundary conditions in mode III}
\label{boundary3}
Far from the crack as $y \rightarrow \pm \infty$ we know $\sigma_{zy} \rightarrow \sigma_{\infty}$ or using the
stress-strain relationships Eq.~(\ref{stress}) we find that $u_z \approx [\sigma_{\infty}/\mu] y$.
Thus the analytic function must have the form 
\begin{equation}
\label{far}
\chi (z) \rightarrow -i [\sigma_{\infty}/\mu] z \quad {\rm as}~ |z| \rightarrow \infty \ .
\end{equation} 

Now on the boundary of the crack the normal stress vanishes, i.e.
\begin{equation}
\label{near}
0=\sigma_{zn}(s) = \partial_n u_z=-\partial_t \xi_z .
\end{equation}
This means that $\xi_z$ is constant on the boundary. We choose 
the gauge $\xi_z=0$, which in
turn is a boundary condition making the analytic function 
$\chi (z)$ real on the boundary of the crack:
\begin{equation}
\label{real}
\chi (z(s)) = \chi (z(s))^* \ .
\end{equation}
\subsection{The stress tensor for mode III}
\label{stress3}

Following the basic strategy we consider now a circular crack in the
mathematical domain.
The strain field for such a crack 
is well known \cite{52Mus}, being the real part of the function $\chi^{(0)}(\omega)$ where
\begin{equation}
\label{solution}
\chi^{(0)} (\omega) = -i [\sigma_{\infty}/\mu](\omega - 1/\omega )
\end{equation}
This is the unique analytic function obeying the boundary conditions
$\chi^{(0)}(\omega) \rightarrow -i[\sigma_{\infty}/\mu]\omega$
as $|\omega| \rightarrow \infty$, while on the unit circle
$\chi^{(0)} (\exp i \theta ) = \chi^{(0)} (\exp i \theta)^*$.
To find the corresponding function in the physical plane is
particularly easy for mode III. Since the real part of the function $\chi(z)$
is analytic, it satisfied Laplace's equation automatically. We only
need to make sure that it satisfies the boundary conditions. However,
if we have a good solution in the mathematical plane, we need just
to compose it with an analytic function that takes us from the
physical to the mathematical plane.
The required analytic function $\chi^{(n)} (z)$ is given by the expression
\begin{equation}
\label{solution2}
\chi^{(n)}  (z) = -i [F_1^{(n)}\sigma_{\infty}/\mu]\Big({\Phi^{(n)}}^{-1}(z) - 1/{\Phi^{(n)}}^{-1}(z)\Big) \ .
\end{equation}
(If $\Phi^{(n)}$ is conformal, ${\Phi^{(n)}}^{-1}$ is analytic by definition).
From this we should compute now
the transverse stress tensor:

\begin{eqnarray}
\label{transverse}
\sigma_{zt}(s) & = & \mu ~\partial_t u_z=\mu~ \Re \frac{\partial\chi^{(n)}(z)}{\partial s}\nonumber\\
&=&\mu ~\Re[\frac{\partial\chi^{(n)}(\Phi^{(n)}(e^{i\theta}))}{\partial \theta} \frac{\partial \theta}
{\partial s}]\nonumber\\&=&-\Re\frac{iF_1^{(n)}\sigma_{\infty} \frac{\partial}{\partial \theta}
(e^{i\theta}-e^{-i\theta})}{|\Phi^{'(n)}(e^{i\theta})|}\nonumber \\
&=&2 \sigma_{\infty}  F_1^{(n)}\frac{ \cos \theta}{|\Phi^{'(n)}(e^{i \theta} )|} \ ,
\end{eqnarray} 
 on the boundary. 
Eqs.(\ref{transverse}) together with (\ref{comp}) 
offer an analytic expression for the transverse stress field at any stage
of the crack propagation.
\begin{figure}
\centering
\includegraphics[width=.3\textwidth]{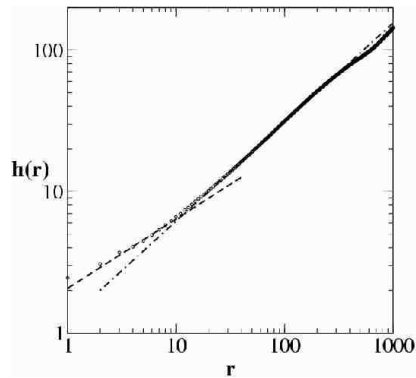}
\caption{$h(r)$ averaged
over all the backbone and over 20 fracture patterns each of which
of 10 000 fracture events. There is a cross-over between a scaling law
with roughness exponent $0.49\pm 0.08$ to an exponent of $0.70\pm 0.05$}
\label{scaling}
\end{figure} 
\section{Results for mode III}
\label{results3}
\subsection{Linear velocity law}
Fig. \ref{fracture} exhibits in the upper panel a typical fracture pattern that is obtained with
this theory, with $\sigma_\infty=1$, after 10 000 growth events. The threshold value
of $\sigma_c$ for the occurrence of the first event (cf. Eq.(\ref{transverse}) is
$\sigma_c=2$. We always implement the first event. For the next growth event 
the threshold is $\sigma_c=2.34315...$.
We thus display in Fig.1 a cluster obtained with
$\sigma_c=2.00$, to be close to the quasi-static limit.
Note that here we could opt to represent a disordered material by
a random value of $\sigma_c$, and see Subsect. \ref{quenched}. With fixed $\sigma_c$, 
one should observe that as the pattern develops,
the stress at the active zone increases, and we get progressively away from
the quasi-static limit. Indeed, as a result of this, for fixed boundary
conditions at infinity, there are more and more values of $\theta$ for which
Eq.(\ref{weight}) does not prohibit growth. Since
the tips of the patterns are mapped by ${\Phi^{(n)}}^{-1}$ to larger and larger
arcs on the unit circle, the support of the probability $P(\theta)$ increases,
and the fracture pattern becomes more and more ramified as the process
advances. The geometric characteristics of the fracture pattern are
{\em not} invariant to the growth. For this reason it makes little sense
to measure the fractal dimension of the pattern; this is not a stable
characteristic, and it will change with the growth. 
On the other hand, we should realize that the fracture pattern is not
what is observed in typical experiments. When the fracture hits the boundaries
of the sample, and the sample breaks into two parts, all the side-branches of the 
pattern remain hidden in the damaged material, and only the backbone
of the fracture pattern appears as the surface of the broken
parts. The backbone does not suffer from the geometric variability discussed above.
In the lower panel of Fig. \ref{fracture} we show the backbone of the pattern displayed
in the upper panel.

This backbone is representative of all the fracture patterns with the
linear velocity law. We should
note that in our theory there are no lateral boundaries, and the
backbone shown does not suffer from finite size effects which may very
well exist in experimental realizations.

In determining the roughness exponent of the backbone, 
we should note that a close examination of it reveals that {\em it is
not a graph}. There are overhangs in this backbone, and since we deal with
mode III fracturing, the two pieces of material {\em can} separate leaving
these overhangs intact. Accordingly, one should not approach the roughness
exponent using correlation function techniques; these may introduce serious
errors when overhangs exist \cite{95OPZ}. Rather, we should measure, for any given $r$,
the quantity \cite{97Bou}
\begin{equation}
h(r) \equiv \langle {\rm Max}\{y(r')\}_{x<r'<x+r}-{\rm Min}\{y(r'\}_{x<r'<x+r}\rangle_x \ .
\end{equation}
The roughness exponent $\zeta$ is then obtained from
\begin{equation}
h(r) \sim r^\zeta  \ ,
\end{equation}
if this relation holds. To get good statistics we average, in addition
to all $x$ for the same backbone,
over many fracture patterns. The result of the analysis
is shown in Fig. \ref{scaling}. 

We find that the roughness exponent for the backbone exhibits a clear
cross-over from about 0.5 for shorter distances $r$ to about 0.70 for larger
distances. Within the error bars these results are in a surprising agreement 
with the numbers quoted experimentally, see for example \cite{97Bou}.
The short length scale exponent of order 0.5 is also in agreement with
recent simulational results of a lattice model \cite{00PCP} (which is by definition
a short length scale solution).
Bouchaud \cite{97Bou} proposed that the cross-over stems from transition between slow
and rapid fracture, from the ``vicinity of the depinning transition" to
the ``moving phase" in her terms. Obviously, in our theory we solve
the quasi-static equation all along, and there is no change of physics.
In addition, there is no reason to expect the experiment to be
a pure mode III, and as we will see below modes I and II do not
show similar roughening.
Nevertheless, as we observed before, the fracture pattern begins with
very low ramification when the stress field exceeds the threshold value
only at few positions on the fracture interface. Later it evolves to a much more ramified
pattern due to the increase of the stress fields at the tips of the
mature pattern. {\em The scaling properties of the backbone reflect
this cross-over}. We propose that this effect is responsible for
the cross-over in the roughening exponent of the backbone. On the
other hand, this non-stationarity in the geometric characteristics
should be handled with care, since it may mean that there is no
definite roughening exponent, as it may depend on {\em where} the
analysis is done, near the center of the fracture patterns or
near the edge. We will return to this delicate issue after reviewing
the results of other velocity laws.
\subsection{Other velocity laws}
\begin{figure}
\centering
\includegraphics[width=.3\textwidth]{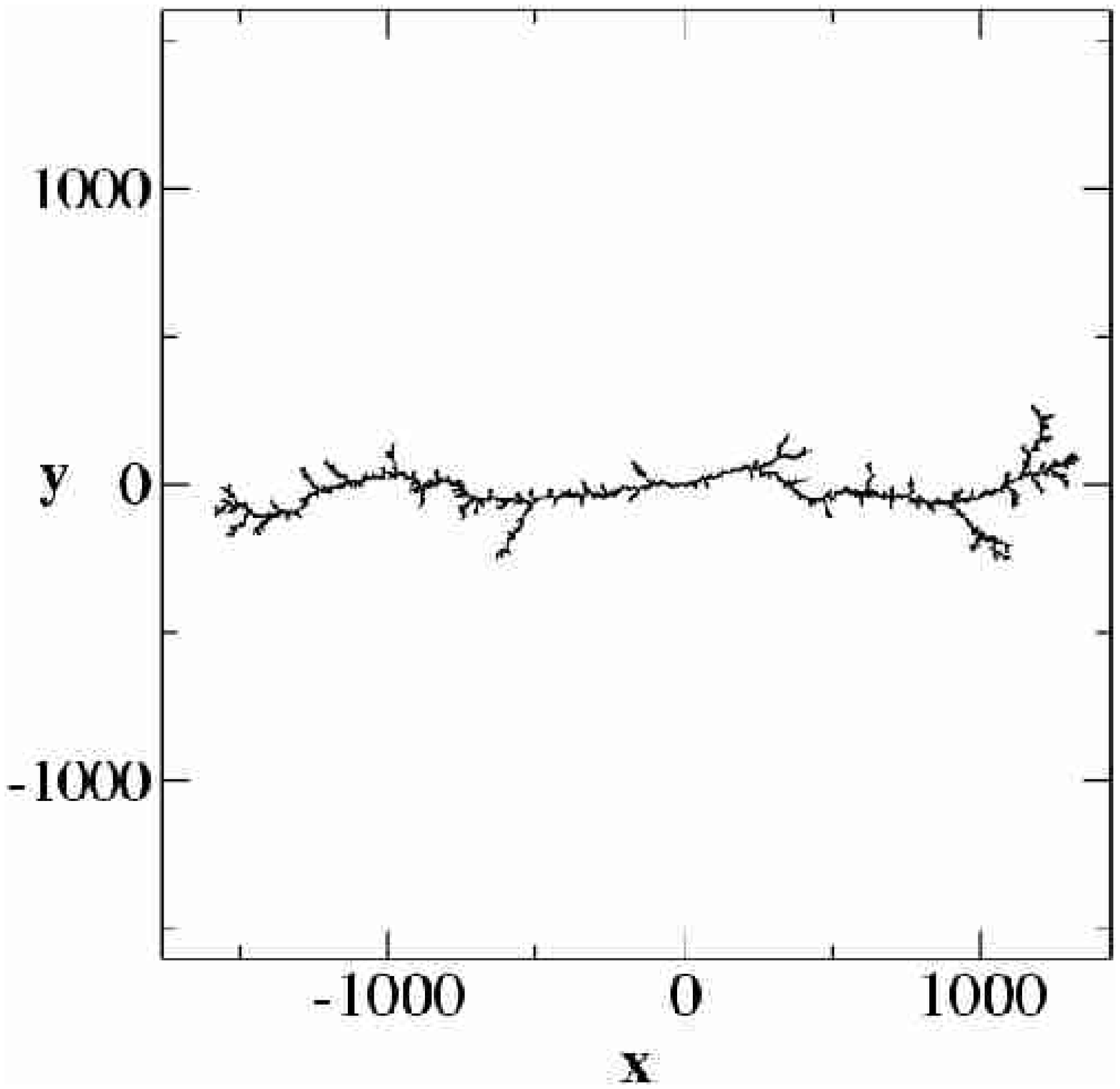}
\includegraphics[width=.3\textwidth]{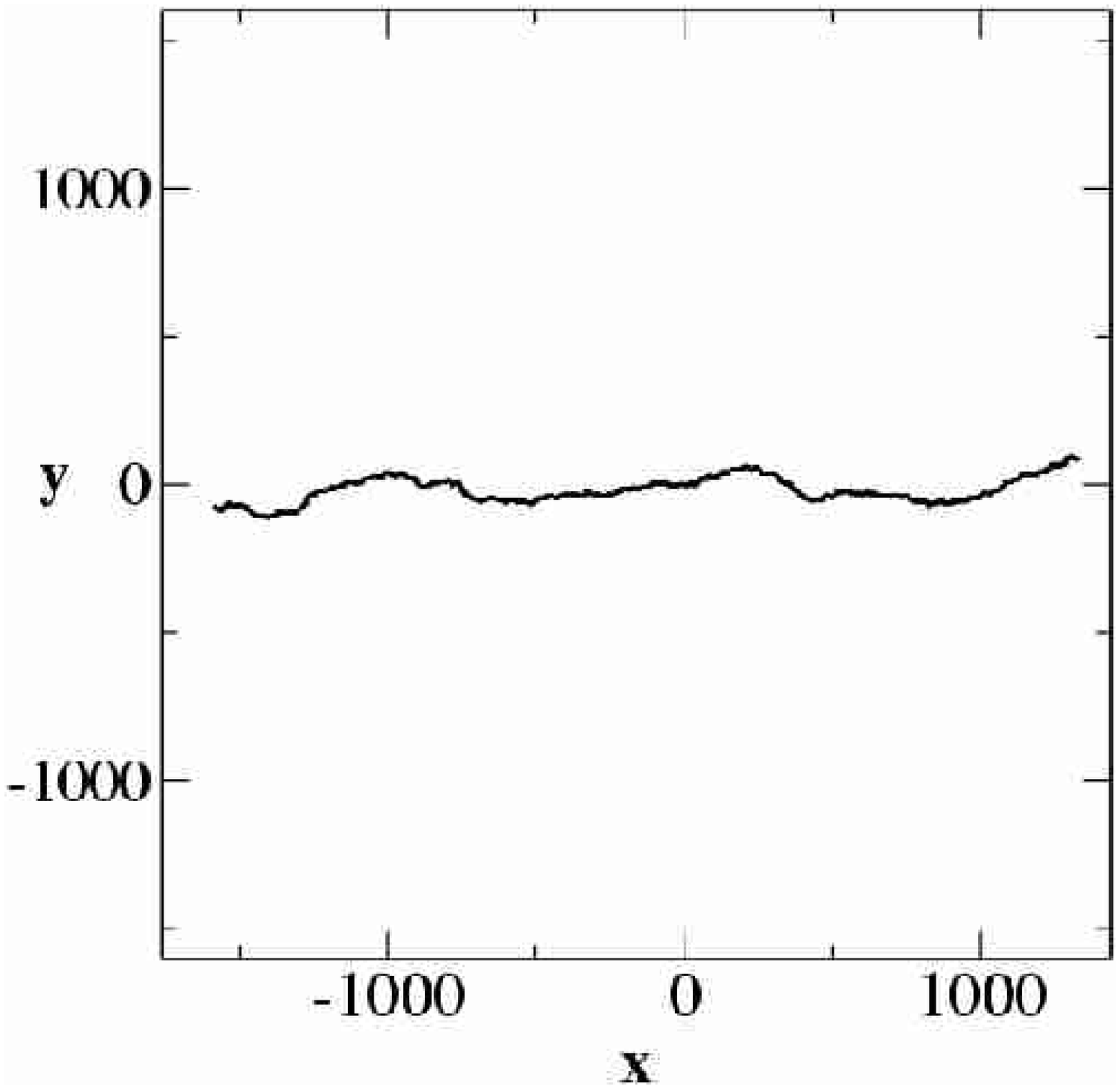}
\caption{Upper panel: fracture pattern for mode III fracture with the quadratic law
(\ref{quad}), with 10 000 fracture events. Lower panel: the backbone of the pattern}
\label{mode3quad}
\end{figure}

\begin{figure}
\centering
\includegraphics[width=.3\textwidth]{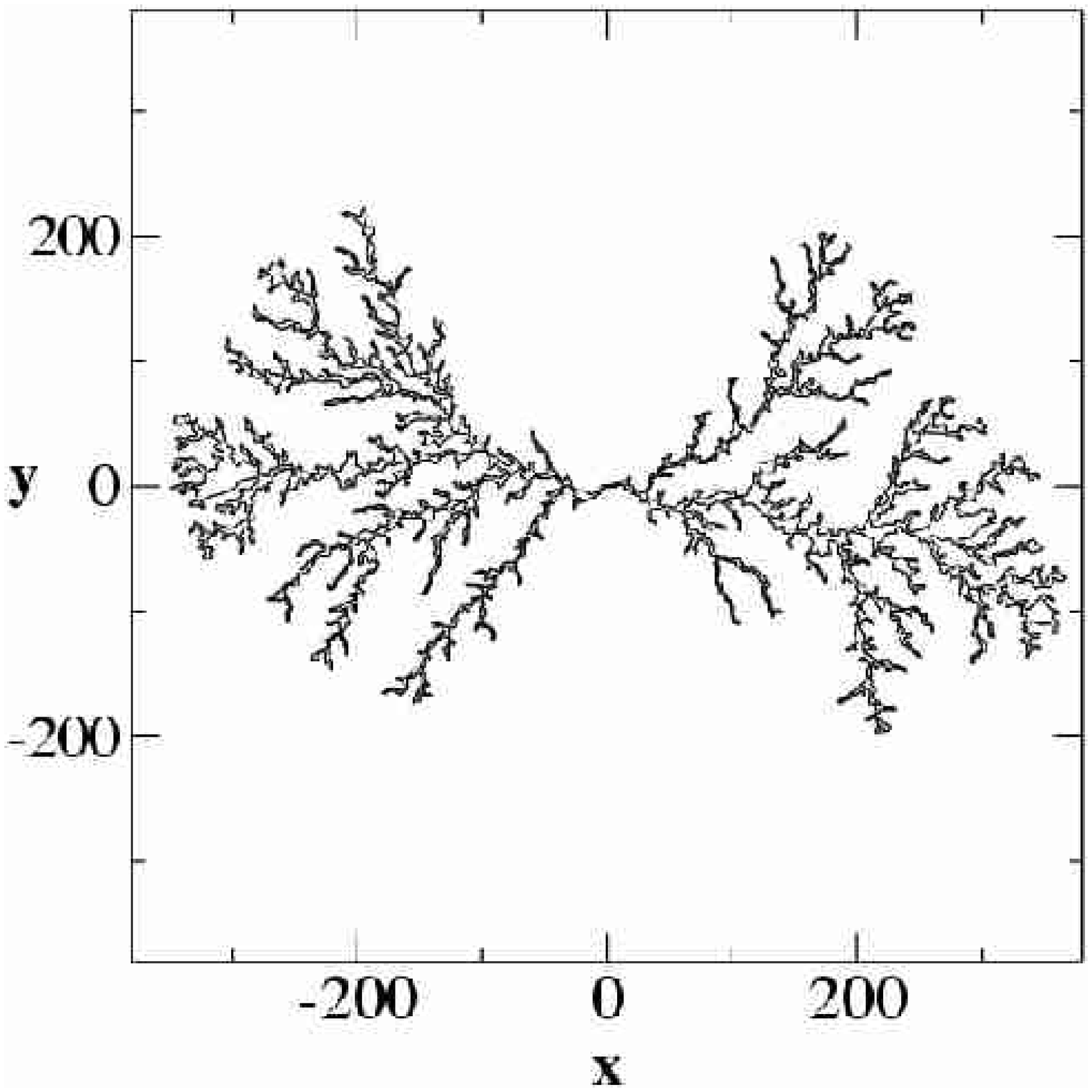}
\includegraphics[width=.3\textwidth]{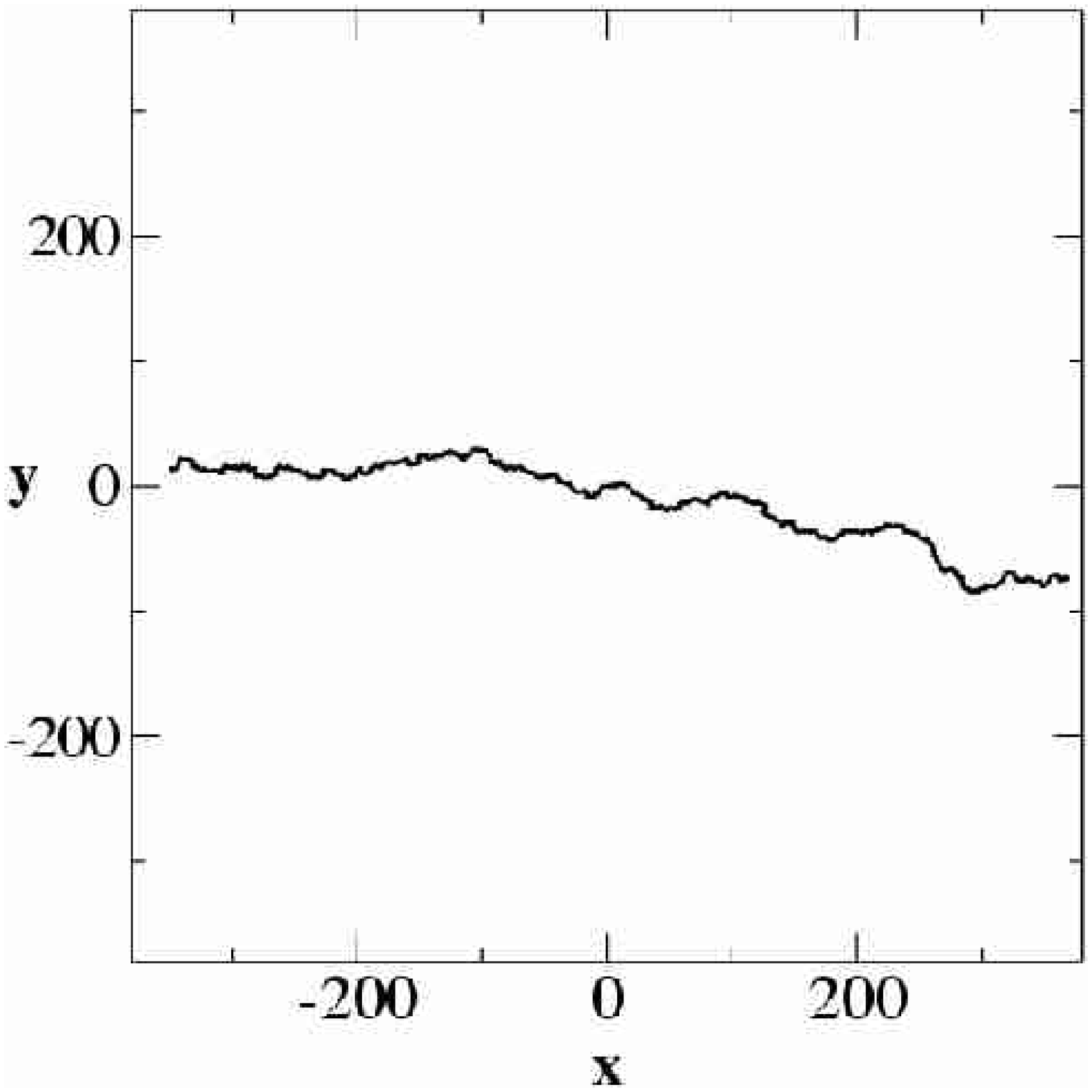}
\caption{Upper panel: fracture pattern for mode III fracture with the exponential law
(\ref{exp}), with $\alpha=0.1$, with 10 000 fracture events. 
Lower panel: the backbone of the pattern}
\label{mode3exp}
\end{figure}

\begin{figure}
\centering
\includegraphics[width=.3\textwidth]{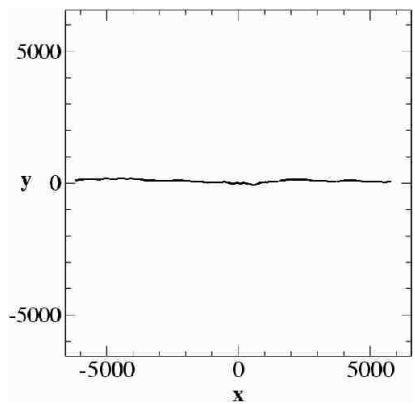}
\caption{ Fracture pattern for mode III fracture with the exponential law
(\ref{exp}), with $\alpha=1$, with 10 000 fracture events. 
In this case the fracture pattern and the
backbone are the same.}
\label{mode3exp2}
\end{figure}

\begin{figure}
\centering
\includegraphics[width=.3\textwidth]{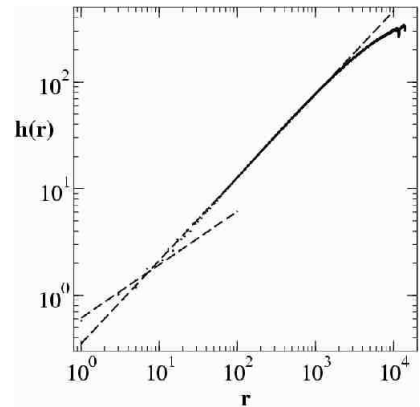}
\caption{ $h(r)$ averaged over
20 fracture patterns with the exponential velocity law with $\alpha=1$. Each of the
patterns consists of 10 000 fracture events. There is a cross-over between a scaling law
with roughness exponent of about $0.50$ at short length scales to
an apparent scaling exponent of about 0.78.}
\label{hrexp2}
\end{figure}

\begin{figure}
\centering
\includegraphics[width=.3\textwidth]{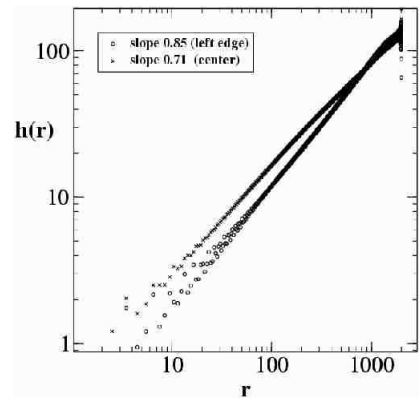}
\caption{ $h(r)$ averaged over
20 fracture patterns with the exponential velocity law with $\alpha=1$. In this
calculation we concentrate on parts of the pattern shown in Fig. \ref{mode3exp2},
one near the center and the other near the edge, each consisting of $r=2000$. 
The apparent exponents differ, being  0.71 at the center and 0.85 near
the edge. The average behavior with exponent 0.78 seen in Fig. \ref{hrexp2}
should be therefore interpreted with extra care.}
\label{2regions}
\end{figure}

It should be stressed that there is no reason to believe that
the scaling exponents are invariant to the change of the velocity
law. In Figs. \ref{mode3quad}, \ref{mode3exp} and \ref{mode3exp2} we
show the fracture patterns and their corresponding backbones
for the quadratic velocity law (\ref{quad})
and for two different exponential laws (\ref{exp}). 
We find that the quadratic
law makes little difference with respect to the linear law. The
roughening plot is similar, and the scaling exponents appear the same. The exponential
velocity law changes the degree of ramification, and therefore
calls for a careful discussion of the roughening plots.  Examine 
the function $h(r)$ for the pattern in
Fig. (\ref{mode3exp2}) (see Fig. \ref{hrexp2}). While the small scale roughening
exponent of about 0.5 is reproduced, it appears that the large scale
exponent is now higher, about 0.78.  
The question to be asked therefore is whether the scaling exponent is not
invariant to the velocity law. In our opinion this question is ill-posed since the scaling
exponent itself {\em depends on where is it measured}. As we said before,
the fracture pattern tends to become more ramified as it grows. This is
reflected in the roughening properties. To make this point clearer, we have
taken the pattern of Fig. \ref{mode3exp2} as a test case, and computed the
apparent scaling exponents for short parts of the fracture pattern, limiting
the maximal value of $r$ to 2000. Doing so, we can concentrate on a region
near the center of the pattern, and on a region near the edge. The results
of this exercise are presented in Fig. \ref{2regions}
What is found is that the apparent scaling exponent depends on the region
of measurements. Near the center, where the pattern is less ramified,
the exponent is smaller than near the edge where the pattern is more
ramified. The average exponent reported in Fig. \ref{hrexp2} which
is analogous to what is reported in experiments, has therefore
a limited value. It may not be interpreted as a 'true' scaling
exponents. Its value may well depend on the actual length of the
pattern that is investigated.

We are therefore not in a
position to claim that the correspondence in roughening exponents between the linear law and experiments
indicates anything about universality classes. One needs to ascertain very carefully whether
measured roughening exponents indicate translationally invariant scaling properties. 
It is in particular useful to know whether the observed scaling exponents depends on the
length of the available fracture pattern. 
\subsection{quenched disorder}
\label{quenched}
To study the effect of quenched randomness we assign a-priori a random
value $\sigma_c$ to every point in the material (with resolution $\lambda_0$).
Not having a clear indication from the literature how the randomness
of inhomogeneous media should be modeled, we opted for two types
of quenched randomness. The first takes the numerical value of
$\sigma_c(s)$ from a flat distribution, $0\le \sigma_c \le \sigma_{max}$
and the second takes a power law form
\begin{equation}
P(\sigma_c) \propto \sigma_c^{-\beta} \ , \quad \text{for}~ \sigma_c>\sigma_{min} \label{Psigma}
\end{equation}

\begin{figure}
\centering
\includegraphics[width=.3\textwidth]{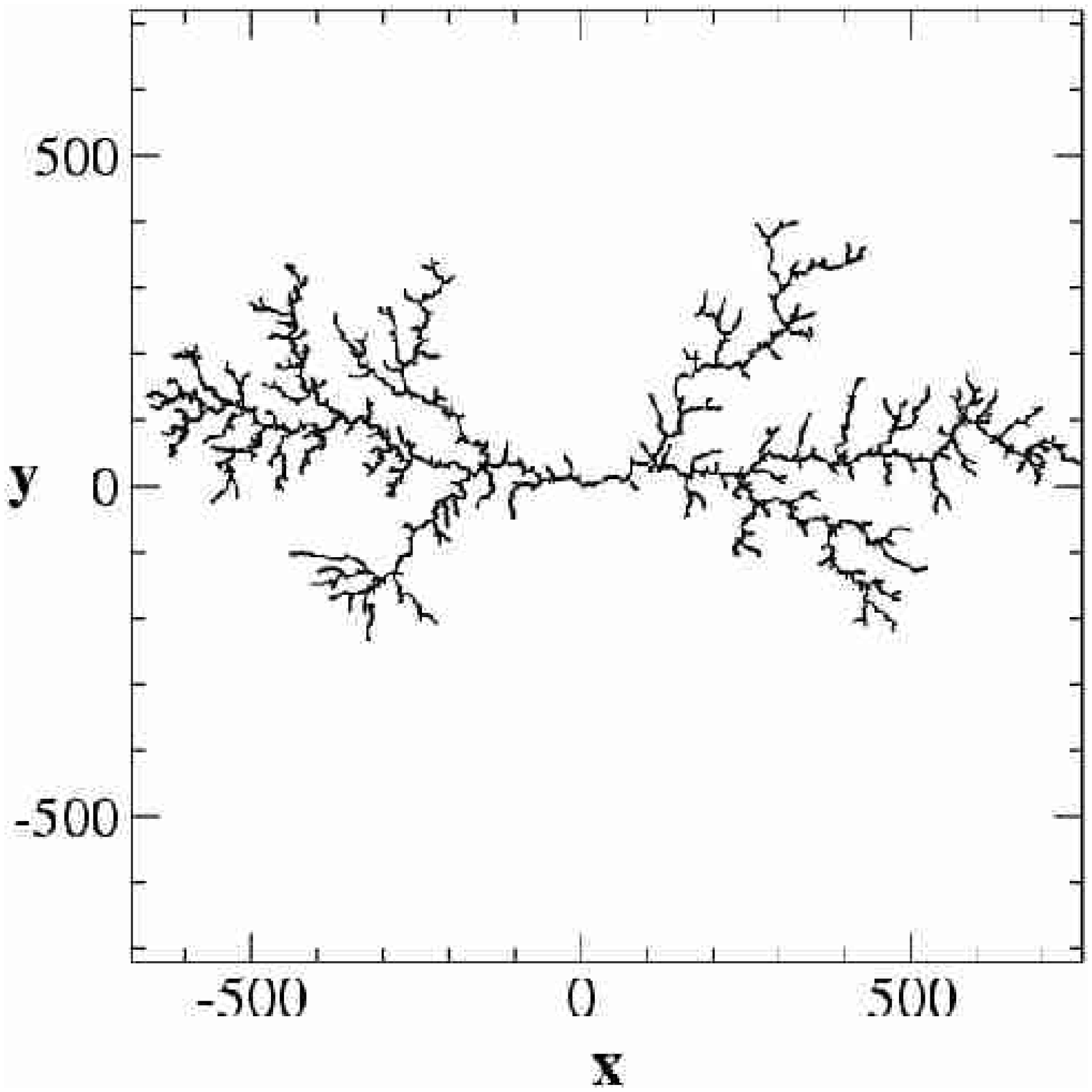}
\includegraphics[width=.3\textwidth]{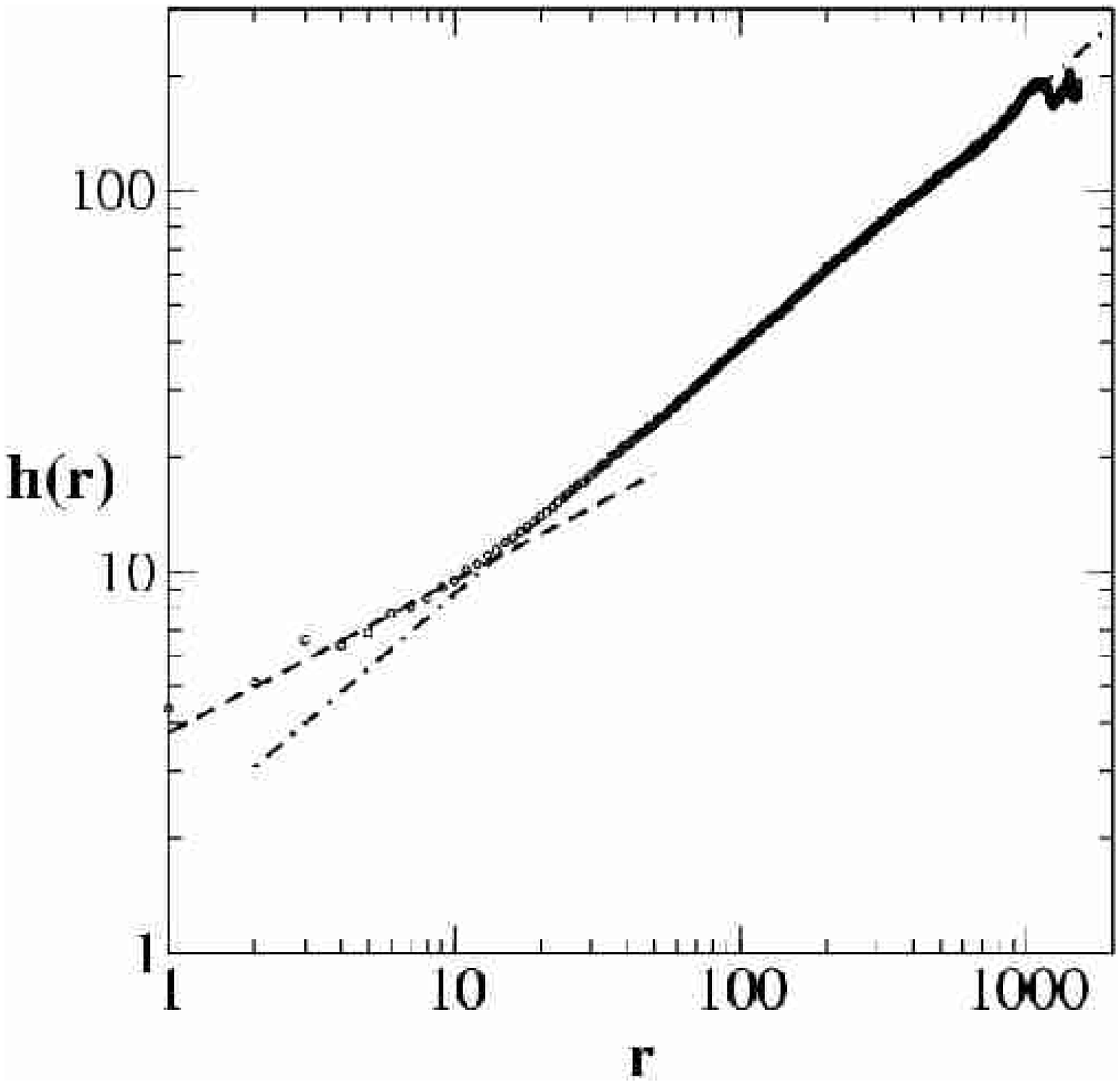}
\caption{Upper panel: fracture pattern for mode III fracture with the linear
velocity law and quenched randomness with a flat distribution, $\sigma_{max}=15$,
with 10 000 fracture events.
Lower panel: the function $h(r)$ after averaging over 20 patterns.
The scaling exponents are about 0.4 and 0.65 for the smaller and
larger scales, respectively.}
\label{quenchflat}
\end{figure}

For reasonable values of $\sigma_{max}$ the flat distribution did not lead
to a qualitative change in the fracture patterns. In Fig. \ref{quenchflat}
we show the pattern and the function $h(r)$ for the case
$\sigma_{max}=15$. 
The typical cross over that we see in systems without quenched disorder remains here, 
albeit with apparently smaller exponents, of about 0.4 and 0.65. 

\begin{figure}
\centering
\includegraphics[width=.3\textwidth]{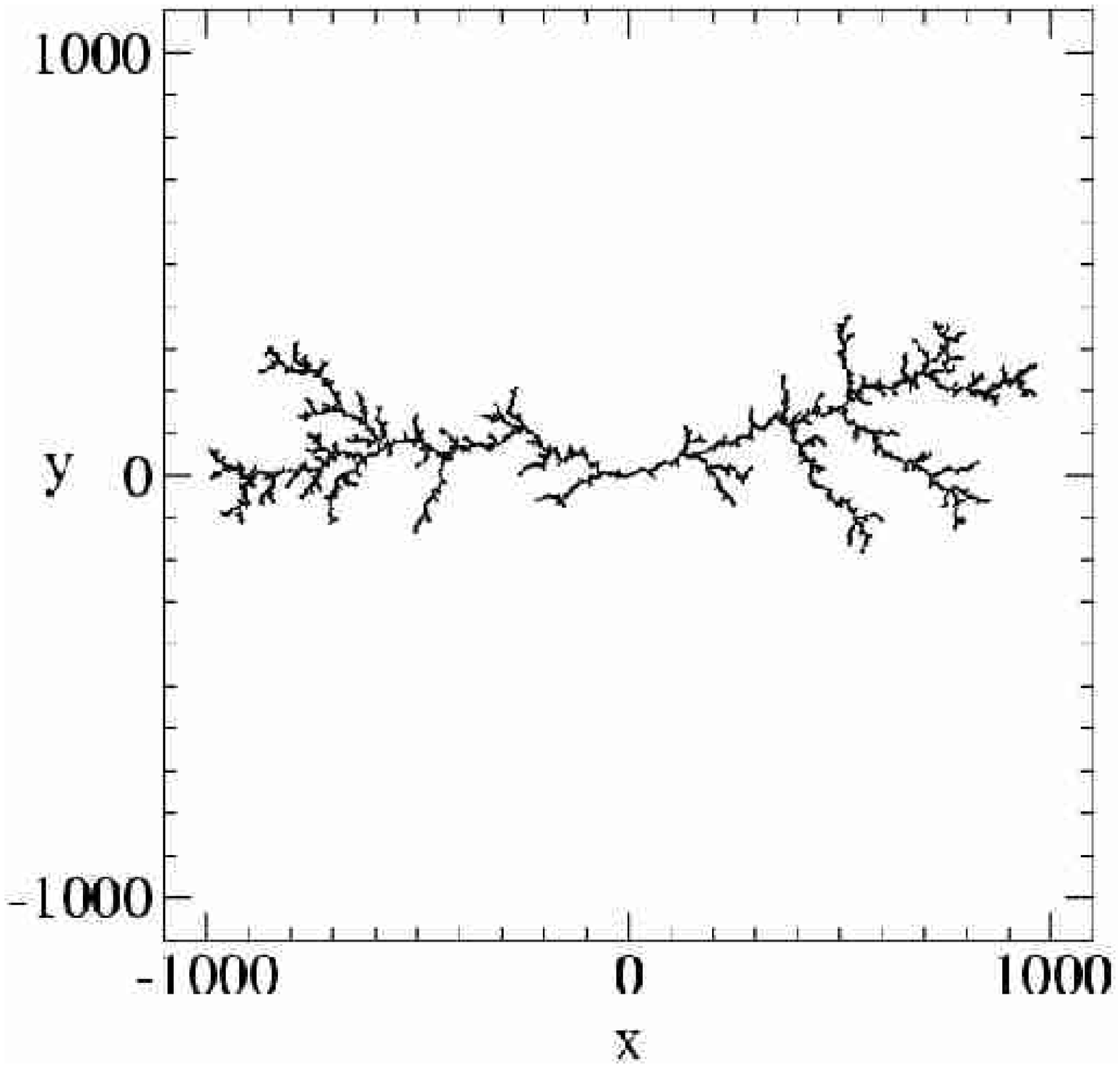}
\includegraphics[width=.3\textwidth]{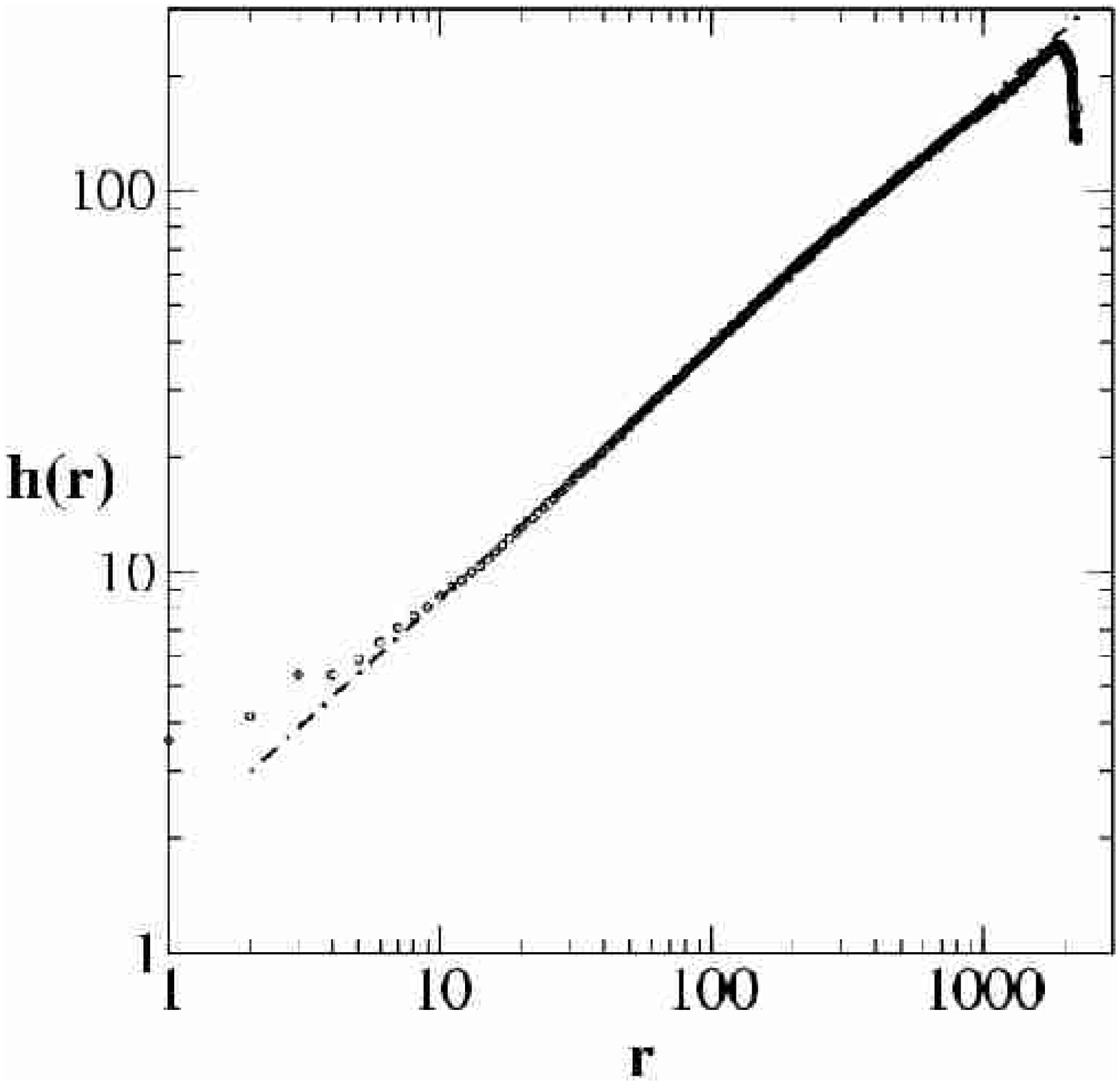}
\caption{Upper panel: Fracture pattern for mode III fracture with the linear
velocity law and quenched randomness with a power-law distribution, $\beta=2, \sigma_{min}=2$,
with 10 000 fracture events. Lower
panel: the function $h(r)$ after averaging over 20 patterns. The scaling exponent is about 0.65.}
\label{quenchpower1}
\end{figure}

On the other hand, a power law distribution of quenched randomness may lead to very 
interesting qualitative change in fracture pattern. While high values
of $\beta$ in (\ref{Psigma}) are still in qualitative agreement with all
previous results (see Fig. \ref{quenchpower1} with $\beta=2$), lower values of $\beta$
lead to a new phenomenon. 
\begin{figure}
\centering
\includegraphics[width=.3\textwidth]{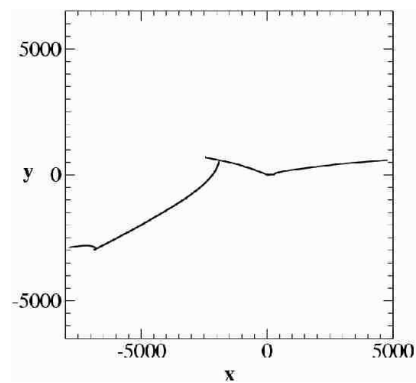}
\caption{Fracture pattern for mode III fracture with the linear
velocity law and quenched randomness with a power-law distribution, $\beta=1.1,\sigma_{min}=0.2$.}
\label{quenchpower2}
\end{figure} 
The availability of very high values
of $\sigma_c$ results in effective blocking for the evolution of the
fracture. The crack develops along continuous (sometime curved)
lines, and then it suddenly gains sharp turns.
In Fig. (\ref{quenchpower2}) we show the typical patterns obtained
for $\beta=1.1$. It is amusing to note that these patterns are reminiscent
of what is exhibited in a number of experiments and see for example
the pictures in \cite{97Bou}. It is not obvious however
how to offer quantitative measures for comparison. It appears to the present
authors that this subject of fracture with quenched randomness deserves
a careful separate study in which experimental and theoretical
methods were combined to gain further insights on the questions
at hand.

\section{Theory for mode I and II}
\label{modeI}

In order to compute the stress tensor at the boundary of the
crack for modes I and II loading, we turn to the solution
 of Eq.(\ref{bilaplace}).
\subsection{Boundary conditions and removal of freedoms}
The boundary conditions
at infinity are given by Eqs. (\ref{mode1}), (\ref{mode2}). The conditions
on the boundary of the crack are
\begin{equation}
\sigma_{xn}(s)=\sigma_{yn}(s)=0  \quad \text{on the boundary}\ . \label{bcm12}
\end{equation}
Using Eq. (\ref{sigU}) these boundary condition are rewritten as
\begin{equation}
\partial_t\left[\frac{\partial U}{\partial x}
+i\frac{\partial U}{\partial y}\right]=0  \quad \text{on the boundary}\ . \label{bcU}
\end{equation}
Note that we do not have enough boundary conditions
to determine $U(x,y)$ uniquely. In fact we can allow in Eq. (\ref{Uphichi}) 
arbitrary transformations of the form 
\begin{equation}
\varphi \rightarrow \varphi +iCz+\gamma
\end{equation}
\begin{equation}
\psi \rightarrow \psi +\tilde\gamma \ , \quad\psi\equiv \eta'
\end{equation}
where $C$ is a real constant and $\gamma$ and $\tilde\gamma$ are 
complex constants. This provides five degrees of freedom in the
definition of the Airy potential. Two of these freedoms are removed
by choosing the gauge in Eq.(\ref{bcU}) according to 
\begin{equation}
\frac{\partial U}{\partial x}
+i\frac{\partial U}{\partial y} = 0  \ , \quad \text{on the boundary}\ . \label{choice}
\end{equation}
It is important to stress that whatever the choice of the five
freedoms the values of the stress tensor are unaffected, and see
\cite{52Mus} for an exhaustive discussion of this point. Computing (\ref{choice}) in 
terms of (\ref{Uphichi}) we arrive at the boundary condition
\begin{equation}
\varphi(z)+z\overline{\varphi'(z)}+\overline{\psi(z)}=0 \text{ on the boundary}
\label{bccrack}
\end{equation}
To proceed we represent $\varphi(z)$ and $\psi(z)$ in
Laurent form:
\begin{eqnarray}
\varphi(z) &=&\varphi_1 z + \varphi_0 
+\varphi_{-1}/z+\varphi_{-2}/z^2+\cdots \ , \nonumber\\
\psi(z) &=&\psi_1 z + \psi_0 
+\psi_{-1}/z+\psi_{-2}/z^2+\cdots \ . \label{Laurentpp}
\end{eqnarray}
This form is in agreement with the boundary conditions at infinity that
disallow higher order terms in $z$. 
The remaining freedoms are now used to choose $\varphi_0 =0$ and $\varphi_1$ real.
Then, using the boundary conditions (\ref{mode1}) and (\ref{mode2}), we find
\begin{eqnarray}
 \varphi_1&=&\frac{\sigma_{\infty}}{4}\ ;\quad 
 \psi_1=\frac{\sigma_{\infty}}{2} \quad  \text{ Mode I}  \ ,\nonumber\\
 \varphi_1&=&0 \ ; \quad\quad \psi_1=i\sigma_\infty  
\quad \text{ Mode II} \ . \label{p1p1}
\end{eqnarray}
\subsection{The Conformal map and its consequences}

The conformal map is identical in form and meaning to the one introduced
above and successfully applied to mode III. On the other hand, at present
we do not solve the Laplace equation, and our fundamental solution
(\ref{Uphichi}) is {\em not} the real part of an analytic function.
We thus cannot simply solve in the mathematical plane and compose
with the inverse of the conformal map.  

In terms of the conformal map we will
write our unknown functions $\varphi(z)$ and $\psi(z)$ as
\begin{equation}
\varphi(z)\equiv \tilde \varphi\left({\Phi^{(n)}}^{-1}(z)\right) \ , \quad 
\psi(z)\equiv \tilde \psi\left({\Phi^{(n)}}^{-1}(z)\right) \ . \label{ppp}
\end{equation}
Using the Laurent form (\ref{Laurent}) of the conformal map the linear
term at $\omega\to \infty$ is determined by Eqs. (\ref{ppp}). We
therefore can write
\begin{eqnarray}
\tilde \varphi(\omega) &=& \varphi_1F_1^{(n)} \omega 
+ \tilde\varphi_{-1}/\omega+\tilde\varphi_{-2}/\omega^2 +\dots\ , \nonumber\\
\tilde \psi(\omega) &=& \psi_1F_1^{(n)} \omega + \tilde \psi_0 +\tilde\psi_{-1}/\omega
+\tilde\psi_{-2}/\omega^2 +\dots
\ . \label{expom}
\end{eqnarray}
The boundary condition (\ref{bccrack}) is now read for the unit circle
in the $\omega$ plane. Denoting $\epsilon\equiv \exp(i\theta)$ and
\begin{equation}
u(\epsilon)\equiv \sum_{n=1}^\infty \tilde\varphi_{-n}/\epsilon^n \ , 
\quad v(\epsilon)\equiv  \sum_{n=0}^\infty \tilde\psi_{-n}/\epsilon^n \ , \label{uvdef}
\end{equation}
we write 
\begin{equation}
u(\epsilon) + \frac{\Phi^{(n)}(\epsilon)}{\overline{\Phi^{'(n)}(\epsilon)}}
\overline{u'(\epsilon)}+\overline{v(\epsilon)}=f(\epsilon) \ .
\label{eq.fund}
\end{equation}
The function $f$ is a known
function that contains all the coefficients that were determined 
so far:
\begin{equation}
f(\epsilon)
=-\varphi_1 F^{(n)}_1\epsilon - \frac{\Phi^{(n)}(\epsilon)}
{\overline{\Phi^{'(n)}(\epsilon)}}\varphi_1 F^{(n)}_1
-\frac{\overline{\psi_1} F^{(n)}_1}{\epsilon}
\label{ff}
\end{equation}
\subsection{Solution by power series}

To solve the problem we need to compute the coefficients $\tilde\varphi_n$ and $\tilde\psi_n$.
To this aim we first represent
\begin{equation}
\frac{\Phi^{(n)}(\epsilon)}{\overline{\Phi^{'(n)}(\epsilon)}}= \sum_{-\infty}^{\infty}b_i\epsilon^i.
\label{expmap}
\end{equation}
The function $f(\sigma)$ has also an expansion of the form
\begin{equation}
f(\epsilon)= \sum_{-\infty}^{\infty}f_{i}\epsilon^i \ . \label{fepsilon}
\end{equation}
In the discussion below we assume that the coefficients $b_i$ and $f_i$ are
known. In fact what is computed in our procedure is the conformal map
$\Phi^{(n)}(\omega)$. Thus to compute these coefficients we need
to Fourier transform the function $\Phi^{(n)}(\epsilon)/
\overline{\Phi^{'(n)}(\epsilon)}$. This is the most expensive step in
our solution, since the branch cuts that exist in Eq. (\ref{phi}) rule
out the use of Fast Fourier Transforms. One needs to carefully evaluate the
Fourier integrals between the branch cuts. The technique how to track the 
position of the branch cuts on the unit circle was developed in \cite{01BDLP}
and \cite{02BDP}; after having the branch cuts the integrals are evaluated over
1000 equi-distant points between each pair of branch cuts.
Using the last two equations together with (\ref{uvdef}) and (\ref{eq.fund})
we get 
\begin{eqnarray}
\tilde \varphi_{-m}&-&\sum_{k=1}^\infty k~b_{-m-k-1} \tilde\varphi^*_{-k} =f_{-m} 
\ , \quad m=1,2\cdots \ , \label{power_sol1}\\
\tilde \psi^*_{-m}&-&\sum_{k=1}^\infty k~b_{m-k-1} \tilde\varphi^*_{-k} =f_{m} 
\ , \quad m=0,1,2\cdots \ \label{power_sol2}
\end{eqnarray}
These sets of linear equations are well posed.  The coefficients 
$\tilde \varphi_{-m}$ can be calculated from equation (\ref{power_sol1}) alone,  and then
they can be used to determine the coefficients $\tilde \psi_{-m}$. This is in fact a proof that
Eq. (\ref{eq.fund}) determines the functions $u$ and $v$ together. This fact
had been proven with some generality in \cite{52Mus}.

For cracks with simple geometry this is all that we need. For example
for a circular crack (a problem that was explicitly solved in \cite{52Mus})
we simply substitute $\Phi^{(n)}(\omega)=\Phi^{(0)}(\omega)=\omega$, and proceed
to solve for $\tilde \varphi$ and $\tilde\psi$, finding finally
\begin{equation}
\tilde \varphi(\omega)=\varphi_1 \omega - \frac{\psi^*_1}{\omega} \ , 
\quad \tilde\psi(\omega)=\psi_1\omega-2\frac{\varphi_1}{\omega}-\frac{\psi_1^*}{\omega^3}
\end{equation}
For developing cracks of arbitrary shape this is just the starting point. As before
in the solution of mode III we need to compute $\sigma_{tt}$ from which we
construct the probability measure for the first fracture event. The development
of the $\Phi^{(n)}$ then follows the same lines as before.

To compute $\sigma_{tt}$ at the boundary of the crack we use the fact that
follows directly from the definitions that
\begin{equation}
\sigma_{xx}+\sigma_{yy} = 4 \text{Re}[\varphi'(z)]=4\text{Re}\left[\frac{\tilde
\varphi'(\omega)}{\Phi^{'(n)}(\omega)}\right] \ .
\end{equation}
Since this is the trace of the stress tensor, which is invariant under
smooth coordinate transformation, it is also equal to $\sigma_{nn}+\sigma_{tt}$.
Using the fact that $\sigma_{nn}$ vanishes on the boundary we can
write finally
\begin{equation}
\sigma_{tt}(\epsilon) = 4\text{Re}\left[\frac{\tilde\varphi'(\epsilon)}{\Phi^{'(n)}(\epsilon)}\right] \ .
\end{equation}
This result is of some importance; it shows that to compute the component $\sigma_{tt}$ 
of the stress tensor {\em
on the boundary} we do not need to compute $\tilde\psi(\epsilon)$ at all. Of course, to know the
stress tensor anywhere else in the body we need both functions. For the growth
algorithm this is not necessary. We note that $\tilde\varphi$ is 
computed from Eqs. (\ref{power_sol1}),
and this contains only $b_{m}$ with negative $m$. 
In order to derive a numerical scheme to compute the tangent stress component $\sigma_{tt}$
on the crack we now truncate the series for $\tilde\varphi$ to get an approximation
\begin{equation}
u(\epsilon)\approx \sum_{n=1}^N \tilde\varphi_{-n}/\epsilon^n \label{uapprox}
\end{equation}
We see from Eq. (\ref{power_sol1}) that if we wish to compute this series up to an order N,
we need to compute the coefficients $b_{-j}$ up to $j\le2N+1$ and then solve
the linear system (\ref{power_sol1}). Note that the approximation in Eq. (\ref{uapprox})
corresponds to a truncation of the series (\ref{expmap}) which in turn corresponds to a 
truncation of the conformal map $\Phi^{(n)}$. Since we are interested in the macroscopic
stress distribution along the fracture rather than in the bumpy microstructure, this effect
is of no harm as long as we choose N large enough to resolve the desired patterns. 
\section{Results for mode I and II}
\label{mode12results}
\subsection{Geometry without quenched disorder}
\begin{figure}
\centering
\includegraphics[width=.3\textwidth]{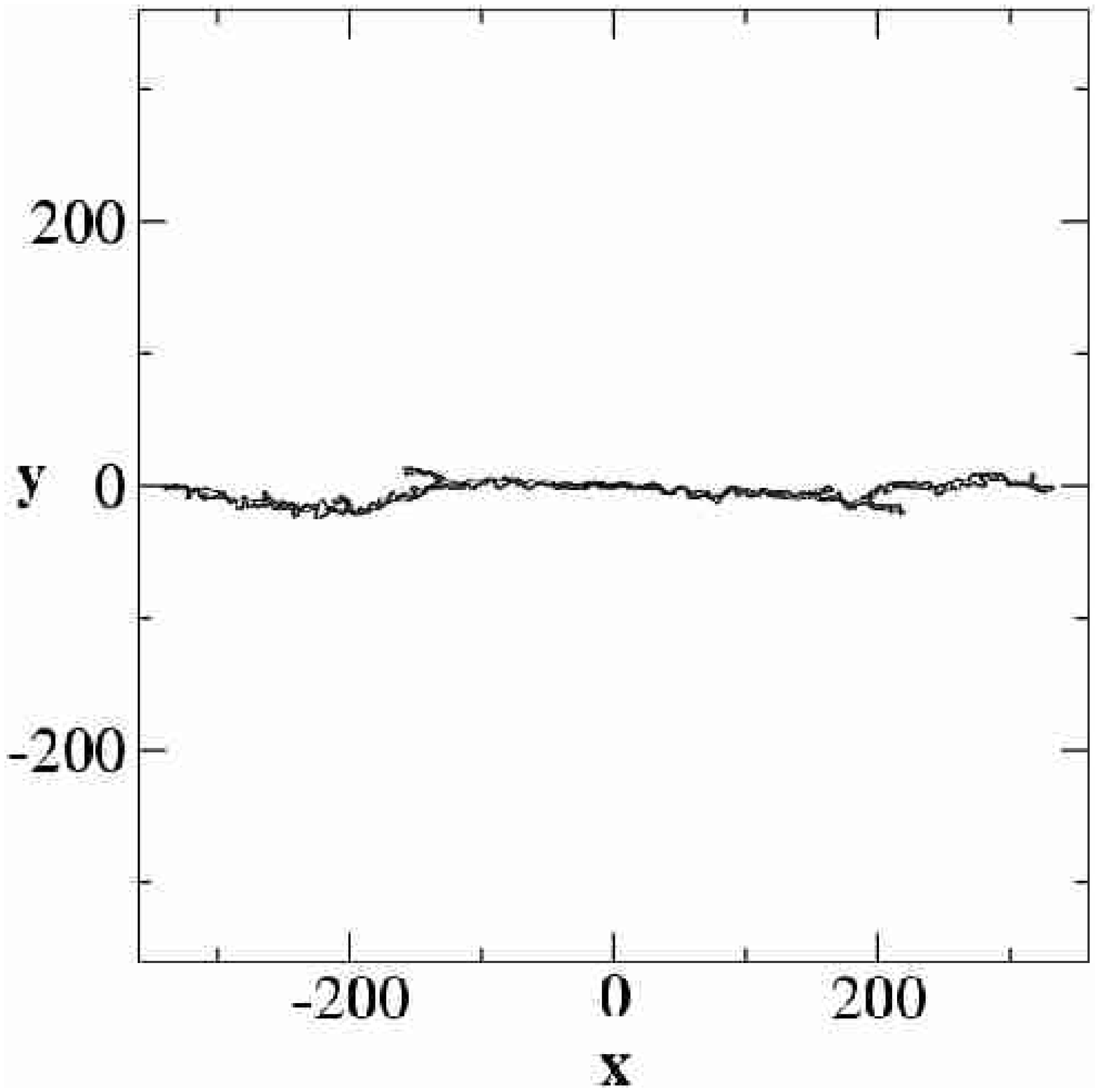}
\includegraphics[width=.3\textwidth]{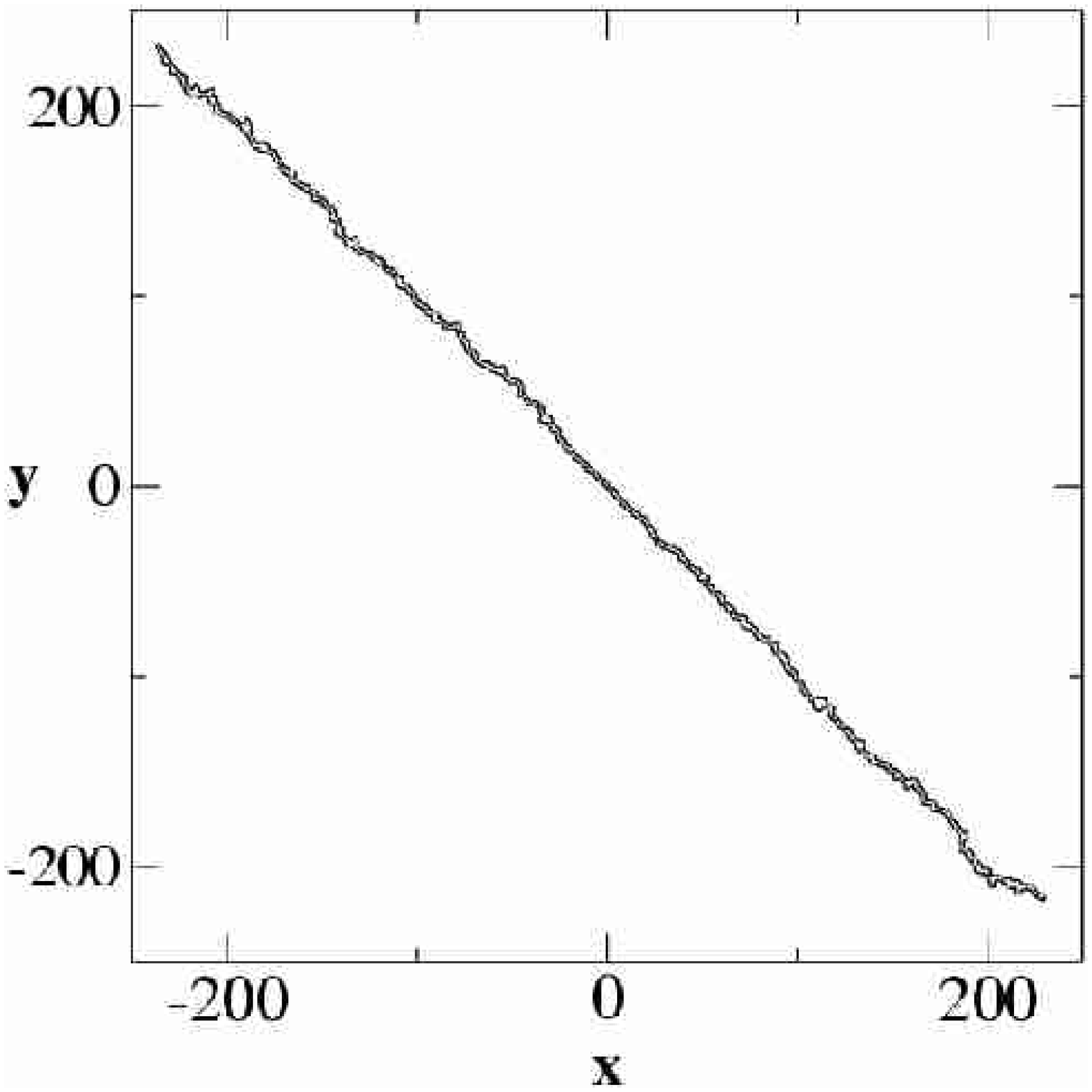}
\caption{Upper panel: fracture pattern for mode I with the linear
velocity law. Lower panel: Fracture pattern for mode II with the linear
velocity law.}
\label{mode12}
\end{figure}

The actual fracture patterns that we find for modes I and II are dramatically
different from those found for mode III for the same velocity law.
In Fig. \ref{mode12} we show the fracture patterns for the linear 
velocity law after about 800 fracture
events. First, mode I and II are very similar, except for the obvious
forty-five degree tilt in mode II due to the tilt of the symmetry axis
of the loading. 
The highly ramified structure seen
in mode III is gone, and the resulting patterns are more akin to the
exponential velocity law in mode III, cf. Fig. \ref{mode3exp2}. The roughening
plot $h(r)$ is also qualitatively different from mode III with the same
velocity law. We do not observe a cross over to a higher exponent, indicating that
there is no increased roughening at large scales. 
\begin{figure}
\centering
\includegraphics[width=.3\textwidth]{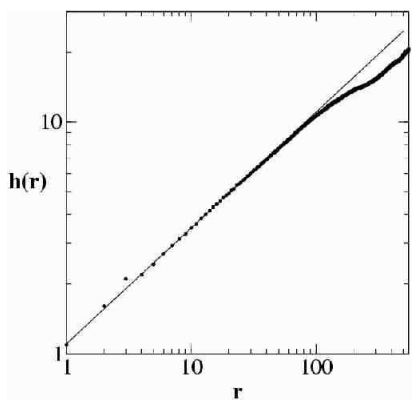}
\caption{The function $h(r)$ for mode I fracture, averaged over
11 fracture patterns. The line indicates a slope of 0.5.}
\label{roughmode1}
\end{figure}
Indeed, for these modes of fracture the stress
field is found to be very highly peaked at the tip of the fracture
pattern. Moreover, when there appear deviations towards side branching
they are quickly corrected in later growth. To make this point clearer
we present in Fig. \ref{sigmadist} the stress field at the boundary of the
crack in the vicinity of the tip. One can observe that the stress component
is such that the slight tilt of the tip will be corrected at the next
growth event. We therefore do not expect large scale roughening in this
mode of fracture.
\begin{figure}
\centering
\includegraphics[width=.3\textwidth]{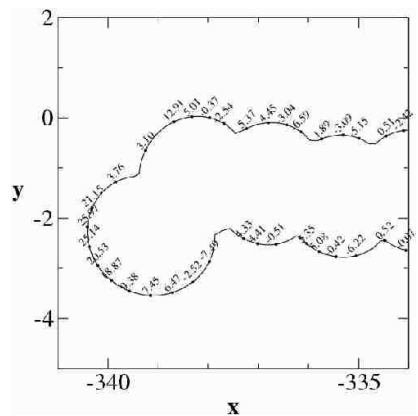}
\caption{The stress field at the boundary of the crack in the vicinity
of the tip.}
\label{sigmadist}
\end{figure}
\subsection{The effect of quenched disorder}

\begin{figure}
\centering
\includegraphics[width=.3\textwidth]{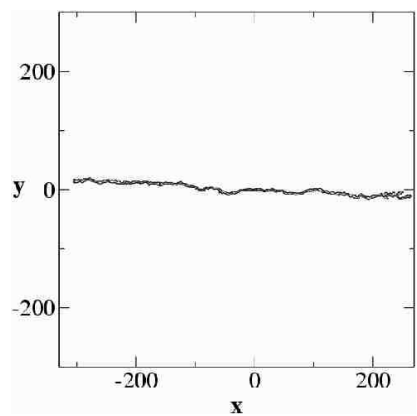}
\caption{The fracture pattern in the case of quenched disorder, with
$\sigma_c$ taken from a flat distribution. The pattern is similar
to the one in Fig. \ref{mode12}, with the same roughening behavior.}
\label{modeIquench}
\end{figure}

Lastly, we present cracks with quenched disorder. First we followed the
growth of a crack in mode I, using the same strategy of Subsect. \ref{quenched}.
In Fig. \ref{modeIquench} show for example the crack obtained with $\sigma_c$
taken from a flat distribution with $\sigma_{max}=10$. Contrary to the 
case of mode III the effect of quenched disorder on the roughening
is not impressive. The roughening exponent is still about 0.5
for small scales, with a failure to roughen on the large scales.
This finding remains invariant to changing the type of quenched disorder
to a power law like Eq.(\ref{Psigma}). We also do not observe 
roughening on the large scales when we put quenched disorder, 
and grow deterministically at the point of highest value
of $\sigma_{tt}-\sigma_c$.
\section{Concluding remarks}
\label{conclusions}
We have presented a solution of the problem quasi-static fracture using
the method of iterated conformal map. All modes of fracture can be
treated, although mode III is much more straightforward since the
equation to be solved is the Laplace equation. The bi-Laplacian equation
that is involved in mode I and II requires heavier analysis and
more cumbersome numerics. Notwithstanding, we believe that our
fracture patterns represent accurate solutions of the problem
with the stated laws of evolution.

The geometric characteristics of mode III are different from
those of modes I and II. The fracture pattern is very ramified,
and if we look at the backbone, (which is what is observed as the
boundary between the two parts of the broken material), we find
that it is rough on all scales. On smaller scales the roughening exponent
is about 0.5, and on larger scales the roughening increases, having
an average roughening exponent which depends on the length of the
fracture pattern analyzed. The exponent 0.5 is intimately
related to the randomness that is introduced by our growth rules.
the higher apparent exponents are due to the increased ramification on the larger
scales as is explained in Sect.\ref{results3}. The roughening
plots may appear to be in close agreement with
some experimental observations, which however are not conducted
as mode III. Experimentally one expects that modes I and II are more
relevant, but here we do not observe the cross over to roughness 
characterized by exponents of the order of 0.75. Quite on the opposite, it appears
that the roughness saturates, leading to a globally flat fracture
patterns on the large scales.

This leaves us with the question of how to interpret the observed
roughness in experiments. One possibility is that experiments are
not quasi-static. We do not have much to say about this possibility.
Another possibility is that in experiments the material has remnant
stresses and other sources of quenched disorder. This is a possibility
that we can put to test. Indeed, we find that mode III is very sensitive
to quenched disorder, cf. Subsect. \ref{quenched}. With power law
disorder we can change the geometric characteristic of the fracture
patterns altogether. This is not the case however with modes I and
II, where the priority of the tip in attracting the stress field
is overwhelming. These cracks do not appear to roughen on the 
large scales even with quenched disorder. 

In summary, we believe that the experimental observations pose
an interesting riddle whose resolution will need a careful assessment
of the experimental conditions and their inclusion in the
theory. It is our hope that the solution presented above will
turn out to be a useful tool in achieving this goal.

\acknowledgments
We thank Jean-Pierre Eckmann for useful discussions on
Fourier transforms in the vicinity of branch-cuts. 
F.B. thanks the "fundacion andes" and the program "inicio de carrera
para jovenes cientificos" C-13760.
This work has been supported in part by the
Petroleum Research Fund, The  European Commission under the
TMR program and the Naftali and Anna
Backenroth-Bronicki Fund for Research in Chaos and Complexity. A. L.
is supported by a fellowship of the Minerva Foundation, Munich, Germany.


\begin{references}
\bibitem{2.3}
See for example \cite{99FM} Sect. 2.3. There is no universal agreement
on the experimental velocity law, and other choices can be made here.
\end{references}
\end{document}